\documentclass{article} % For LaTeX2e
\usepackage{lmodern}
\usepackage[T1]{fontenc}
\usepackage{iclr2025_conference,times}

\usepackage{amsmath,amsfonts,bm}

\def\eqref#1{equation~\ref{#1}}
\def\1{\bm{1}}

\DeclareMathAlphabet{\mathsfit}{\encodingdefault}{\sfdefault}{m}{sl}
\SetMathAlphabet{\mathsfit}{bold}{\encodingdefault}{\sfdefault}{bx}{n}

\usepackage{hyperref}
\hypersetup{
	colorlinks=true,
	linkcolor=red,
	filecolor=blue,
	urlcolor=magenta,
	citecolor=black,
}
\usepackage{url}
\usepackage[flushleft]{threeparttable}
\usepackage{booktabs}

\usepackage{bbding}
\usepackage{tablefootnote}
\usepackage[misc]{ifsym}
\usepackage{colortbl}
\usepackage{multicol}
\usepackage{color}
\usepackage{tabulary}
\usepackage{pifont}
\usepackage{etoolbox}
\definecolor{mycolor_blue}{HTML}{E7EFFA}
\definecolor{mycolor_green}{HTML}{E6F8E0}
\definecolor{mycolor_gray}{HTML}{ECECEC}
\definecolor{pearDark}{HTML}{2980B9}
\usepackage{lipsum}
\usepackage{amssymb}
\usepackage{graphicx}
\usepackage{tcolorbox}
\tcbuselibrary{skins}
\tcbuselibrary{listingsutf8}
\usepackage{multirow}
\usepackage{lipsum}
\usepackage[capitalize,noabbrev]{cleveref}
\usepackage{amsmath}
\usepackage{amssymb}
\usepackage{enumitem}
\usepackage{geometry}
\usepackage{xspace}

\definecolor{stepcolor}{RGB}{54, 162, 235}          % Blue
\definecolor{keycolor}{RGB}{255, 159, 64}          % Orange
\definecolor{generalizedcolor}{RGB}{75, 192, 192}  % Teal
\definecolor{answercolor}{RGB}{153, 102, 255}      % Purple
\tcbset{
    mybox/.style={
        enhanced,
        colback=white,
        colframe=gray!50,
        boxrule=1pt,
        arc=4mm,
        auto outer arc,
        fonttitle=\bfseries,
        title={Solution Steps},
        drop shadow southeast,
        width=\textwidth,
        before skip=10pt,
        after skip=10pt
    }
}

\definecolor{BlueLUH}{cmyk}{1.0,0.7,0,0}
\colorlet{LightBlue}{BlueLUH!20!white}
\colorlet{DarkBlue}{BlueLUH!80!black!20}
\colorlet{PinKish}{red!60}
\definecolor{commentsColor}{rgb}{0.497495, 0.497587, 0.497464}
\definecolor{keywordsColor}{rgb}{0.000000, 0.000000, 0.635294}
\definecolor{stringColor}{rgb}{0.558215, 0.000000, 0.135316}
\definecolor{codegreen}{rgb}{0,0.6,0}
\definecolor{codegray}{rgb}{0.5,0.5,0.5}
\definecolor{codepurple}{rgb}{0.58,0,0.82}
\definecolor{backcolour}{rgb}{0.95,0.95,0.92}
\definecolor{codeblue}{rgb}{0.1, 0.1, 0.8}

\usepackage{tcolorbox}
\usepackage{wrapfig}
\usetikzlibrary{decorations.pathreplacing}
\usetikzlibrary{graphs}
\usepackage{tikz}
\usepackage{pgfplots}
\usepackage{array, threeparttable, booktabs}
\usepackage{algorithm,algpseudocode}
\usepackage{amsthm}

\definecolor{mybrown}{RGB}{128,64,0}
\gdef\Sepline{%
	\par\noindent
	\tikz\draw[thick,dashed,gray!60] (0,0) -- (\linewidth,0);
	\par\nobreak}

\usepackage{tikz}
\usetikzlibrary{quantikz}
\usepackage{tikz-cd}
\tikzcdset{scale cd/.style={every label/.append style={scale=#1},
		cells={nodes={scale=#1}}}}
\usepackage{physics}
\usepackage[outline]{contour} % glow around text
\usetikzlibrary{intersections}
\usetikzlibrary{decorations.markings}
\usetikzlibrary{angles,quotes} % for pic
\usetikzlibrary{calc}
\usetikzlibrary{3d}

\tikzset{>=latex} % for LaTeX arrow head
\colorlet{myred}{red!85!black}
\colorlet{myblue}{blue!80!black}
\colorlet{mycyan}{cyan!80!black}
\colorlet{mygreen}{green!70!black}
\colorlet{myorange}{orange!90!black!80}
\colorlet{mypurple}{red!50!blue!90!black!80}
\colorlet{mydarkred}{myred!80!black}
\colorlet{mydarkblue}{myblue!80!black}
\tikzstyle{xline}=[myblue,thick]

\tikzstyle{myarr}=[myblue!50,-{Latex[length=3,width=2]}]

\usepackage{lipsum}
\usepackage{adjustbox}
\usepackage{longtable}
\usepackage{colortbl}
\usetikzlibrary{arrows}
\usepackage{makeidx}\makeindex
\tikzset{
	operator/.append style={fill=purple!20},
	my label/.append style={above right,xshift=0.3cm},
	phase label/.append style={label position=above}
}

\usepackage{silence}
\definecolor{arsenic}{rgb}{0.23, 0.27, 0.29}
\definecolor{fluxcolor}{RGB}{204, 217, 255}
\definecolor{uwavecolor}{RGB}{244, 220, 222}
\definecolor{FandUwavecolor}{RGB}{231,244,224}
\definecolor{cavitycolor}{RGB}{232, 200, 244}
\definecolor{orangeb}{rgb}{0.99,0.78,0.07}
\definecolor{orangebdark}{rgb}{0.99,0.78,0.17}
\definecolor{livingcoral}{HTML}{FA7268}			% pantone color of the year 2019: 16-1546 Living Coral
\definecolor{ultraviolet}{HTML}{5F4B8B}			% pantone color of the year 2018: 18-3838 Ultra Violet
\definecolor{greenery}{HTML}{88B04B}			% pantone color of the year 2017: 15-0343 Greenery
\definecolor{radiantorchid}{HTML}{AD5E99}		% pantone color of the year 2014: 18-3224 Radiant Orchid
\definecolor{tangerinetango}{HTML}{DD4124}		% pantone color of the year 2011: 17-1463 Tangerine Tango

\DefineNamedColor{named}{Salmon}        {cmyk}{0,0.53,0.38,0}
\DefineNamedColor{named}{CarnationPink} {cmyk}{0,0.63,0,0}

\usepackage{float}

\definecolor{quantumviolet}{HTML}{53257F} %Quantum violet
\definecolor{quantumgray}{HTML}{A9A9A9} %Quantum gray
\definecolor{mygray}{gray}{0.95} %Quantum gray

\newtcolorbox[auto counter,number within=section]{boxfigure}[2][]{%
	colback=mygray,colframe=quantumviolet,fonttitle=\bfseries,width=\textwidth,float*=ht,lower separated=false, halign=justify,title=Box~\thetcbcounter: #2,#1}

\usepackage{listings} % code listings
\definecolor{commentsColor}{rgb}{0.497495, 0.497587, 0.497464}
\definecolor{keywordsColor}{rgb}{0.000000, 0.000000, 0.635294}
\definecolor{stringColor}{rgb}{0.558215, 0.000000, 0.135316}

\usepackage{caption}
\definecolor{codegreen}{rgb}{0,0.6,0}
\definecolor{codegray}{rgb}{0.5,0.5,0.5}
\definecolor{codepurple}{rgb}{0.58,0,0.82}
\definecolor{backcolour}{rgb}{0.95,0.95,0.92}
\definecolor{codeblue}{rgb}{0.1, 0.1, 0.8}

\usepackage{multicol}

\lstdefinestyle{llmprompt001}{
	basicstyle=\footnotesize,        % the size of the fonts that are used for the code
	backgroundcolor=\color{backcolour},
	commentstyle=\color{codegreen},
	keywordstyle=\color{magenta},
	numberstyle=\tiny\color{codegray},
	stringstyle=\color{codepurple},
	basicstyle=\ttfamily\footnotesize,
	breakatwhitespace=false,
	breaklines=true,
	captionpos=b,
	keepspaces=true,
	numbers=left,
	numbersep=5pt,
	showspaces=false,
	showstringspaces=false,
	showtabs=false,
	tabsize=2,
	keywordstyle=\color{codeblue},
	commentstyle=\color{codegreen},
}

\lstdefinestyle{llmprompt002}{
	basicstyle=\footnotesize,        % the size of the fonts that are used for the code
	breakatwhitespace=false,         % sets if automatic breaks should only happen at whitespace
	breaklines=true,                 % sets automatic line breaking
	captionpos=b,                    % sets the caption-position to bottom
	commentstyle=\color{commentsColor}\textit,    % comment style
	deletekeywords={...},            % if you want to delete keywords from the given language
	escapeinside={\%*}{*)},          % if you want to add LaTeX within your code
	extendedchars=true,              % lets you use non-ASCII characters; for 8-bits encodings only, does not work with UTF-8
	frame=tb,	                   	   % adds a frame around the code
	keepspaces=true,                 % keeps spaces in text, useful for keeping indentation of code (possibly needs columns=flexible)
	keywordstyle=\color{keywordsColor}\bfseries,       % keyword style
	language=C++,                 % the language of the code (can be overrided per snippet)
	otherkeywords={*,ket,bra,cmat},           % if you want to add more keywords to the set
	numbers=left,                    % where to put the line-numbers; possible values are (none, left, right)
	numbersep=0pt,                   % how far the line-numbers are from the code
	numberstyle=\tiny\color{commentsColor}, % the style that is used for the line-numbers
	rulecolor=\color{black},         % if not set, the frame-color may be changed on line-breaks within not-black text (e.g. comments (green here))
	showspaces=false,                % show spaces everywhere adding particular underscores; it overrides 'showstringspaces'
	showstringspaces=false,          % underline spaces within strings only
	showtabs=false,                  % show tabs within strings adding particular underscores
	stepnumber=1,                    % the step between two line-numbers. If it's 1, each line will be numbered
	stringstyle=\color{stringColor}, % string literal style
	tabsize=1,	                   % sets default tabsize to 2 spaces
	title=\lstname,                  % show the filename of files included with \lstinputlisting; also try caption instead of title
	columns=fixed     ,               % Using fixed column width (for e.g. nice alignment)
	columns=fullflexible,
	breaklines=true,
	breakatwhitespace=true,
	extendedchars=true,
	breaklines=true,
	xleftmargin=.02\textwidth, % experimentally determined . should fix
	columns=fullflexible,
	flexiblecolumns=true,
}

\definecolor{carrotorange}{rgb}{0.93, 0.57, 0.13}
\usepackage{tabularx}
\usepackage{ltablex}

\DeclareCaptionLabelFormat{boxed}{%
	\kern0.01em{\color{carrotorange}\rule{0.73em}{0.73em}}%
	\hspace*{0.47em}\bothIfFirst{#1}{~}#2}
\usepackage{svg}
\usepackage{amssymb}
\usepackage{pifont}
\title{QuantumLLMInstruct: A 500k LLM Instruction-Tuning Dataset with Problem-Solution Pairs for Quantum Computing.}
\author{\raisebox{-0.2em}{\textcolor{DarkBlue}{\rule{0.8em}{0.8em}}} Shlomo Kashani \textsuperscript{\Letter}\url{skashan2@alumni.jh.edu}\\
	\raisebox{-0.2em}{\textcolor{codeblue}{\rule{0.8em}{0.8em}}} Johns Hopkins University. \\
	\raisebox{-0.2em}{\textcolor{backcolour}{\rule{0.8em}{0.8em}}} Code: \href{https://huggingface.co/spaces/BoltzmannEntropy/QuantumLLMInstruct}{QuantumLLMInstruct HF Code} \\
	\raisebox{-0.2em}{\textcolor{codeblue}{\rule{0.8em}{0.8em}}} Dataset: \href{https://huggingface.co/datasets/BoltzmannEntropy/QuantumLLMInstruct}{QuantumLLMInstruct HF DB} \\
		\raisebox{-0.2em}{\textcolor{backcolour}{\rule{0.8em}{0.8em}}} Project: \href{https://diffusion-book.com/QuantumLLMInstruct.github.io/}{QuantumLLMInstruct Project}
}

\iclrfinalcopy % Uncomment for camera-ready version, but NOT for submission.

\usepackage[colorlinks]{hyperref}
\AtBeginDocument{%
	\hypersetup{
		citecolor=violet,
		linkcolor=violet,
		urlcolor=violet}}

\theoremstyle{definition}

\theoremstyle{plain}

\theoremstyle{definition}
\theoremstyle{plain}

\begin{document}
\maketitle

\begin{abstract}
	We present \textbf{QuantumLLMInstruct (QLMMI)}, an innovative dataset featuring over 500,000 meticulously curated instruction-following problem-solution pairs designed specifically for quantum computing -- \textbf{the largest and most comprehensive dataset of its kind}. Originating from over 90 primary seed domains and encompassing hundreds of subdomains autonomously generated by LLMs, QLMMI marks a transformative step in the diversity and richness of quantum computing datasets.

	Designed for \textbf{instruction fine-tuning}, QLMMI seeks to significantly improve LLM performance in addressing complex quantum computing challenges across a wide range of quantum physics topics. While Large Language Models (LLMs) have propelled advancements in computational science with datasets like \textbf{Omni-MATH} and \textbf{OpenMathInstruct}, these primarily target Olympiad-level mathematics, leaving quantum computing largely unexplored.

	The creation of QLMMI follows a rigorous \textbf{four-stage methodology}. Initially, foundational problems are developed using predefined templates, focusing on critical areas such as synthetic Hamiltonians, QASM code generation, Jordan-Wigner transformations, and Trotter-Suzuki quantum circuit decompositions. Next, detailed and domain-specific solutions are crafted to ensure accuracy and relevance. In the third stage, the dataset is enriched through advanced reasoning techniques, including Chain-of-Thought (CoT) and Task-Oriented Reasoning and Action (ToRA), which enhance problem-solution diversity while adhering to strict mathematical standards. Lastly, a zero-shot Judge LLM performs self-assessments to validate the dataset's quality and reliability, minimizing human oversight requirements.

	To foster collaboration in this evolving field, we utilize the \textbf{Qwen-2.5-Coder} family of models, selected for their strong mathematical reasoning capabilities and permissive licensing. \textbf{All associated code and the complete dataset are openly available}, providing a foundation for future breakthroughs in quantum computing applications of LLMs. Our pipeline is generic and can be easily adapted to generate instruction-tuning datasets across any domain. This work focuses exclusively on the \textbf{generation of synthetic data for quantum computing tasks, deliberately excluding the training or fine-tuning of large LLMs}. No trained model is provided in this study. Instead, we prioritize creating a high-quality dataset to support future research without the computational demands of training or fine-tuning.

\end{abstract}

\setcounter{tocdepth}{4}
\tableofcontents

\section{Introduction}

\begin{figure}[h!]
	\centering
	\includegraphics[width=\textwidth]{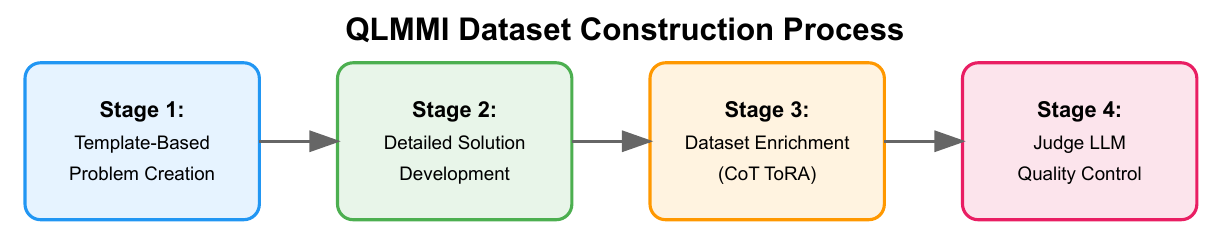}
	\caption{Dataset Creation Workflow for QuantumLLMInstruct. The process comprises four stages: \textbf{Stage I} generates foundational problems with LLMs, focusing on tasks such as synthetic Hamiltonians, QASM code, Jordan-Wigner transformations, and Trotter-Suzuki decompositions for models like Ising~\cite{Ising1925} and Heisenberg spin chains~\cite{Heisenberg1928}. \textbf{Stage II} develops solutions to these problems using LLMs. \textbf{Stage III} enriches the dataset using CoT~\cite{wei2022chain} and ToRA~\cite{wu2023tora}. \textbf{Stage IV} validates the dataset through self-critique mechanisms~\cite{madaan2024self} performed by a Judge LLM, ensuring high-quality entries.}
	\label{fig:db-flow}
\end{figure}

The rapid advancement of Large Language Models (LLMs) has catalyzed transformative breakthroughs across diverse fields, including mathematics~\cite{wei2022chain, brown2020language, wei2022emergent} and natural language processing. Notable models such as GPT~\cite{brown2020language, wei2022cot}, Qwen~\cite{yang2024qwen2, qwen2024}, and LLaMA~\cite{touvron2023llama, touvron2023llama2} have demonstrated exceptional capabilities in solving complex mathematical problems using advanced reasoning methodologies like Chain-of-Thought (CoT)~\cite{wei2022chain, wang2022selfconsistency} and Task-Oriented Reasoning and Action (ToRA)~\cite{wu2023tora}. However, their potential in quantum computing—a domain characterized by intricate reasoning complexity and abstract mathematical frameworks—remains relatively untapped.

\paragraph{Quantum Physics Domains}

Quantum computing tasks, such as simulating time evolution under Hamiltonians~\cite{Trotter1959, Suzuki1976}, compressing quantum models with Yang-Baxter equations~\cite{Peruzzo2014, Heisenberg1928}, or designing circuits for advanced algorithms~\cite{Lloyd1996, Cross2017}, demand a combination of computational rigor and domain expertise. These tasks involve key quantum concepts such as Hamiltonians, entangled states, and unitary transformations, making manual approaches error-prone, resource-intensive, and challenging to scale. The increasing complexity and demand for scalability highlight the critical need for autonomous systems capable of generating, solving, and validating quantum problems at scale.

\begin{figure}[h!]
	\centering
	\includegraphics[width=\textwidth]{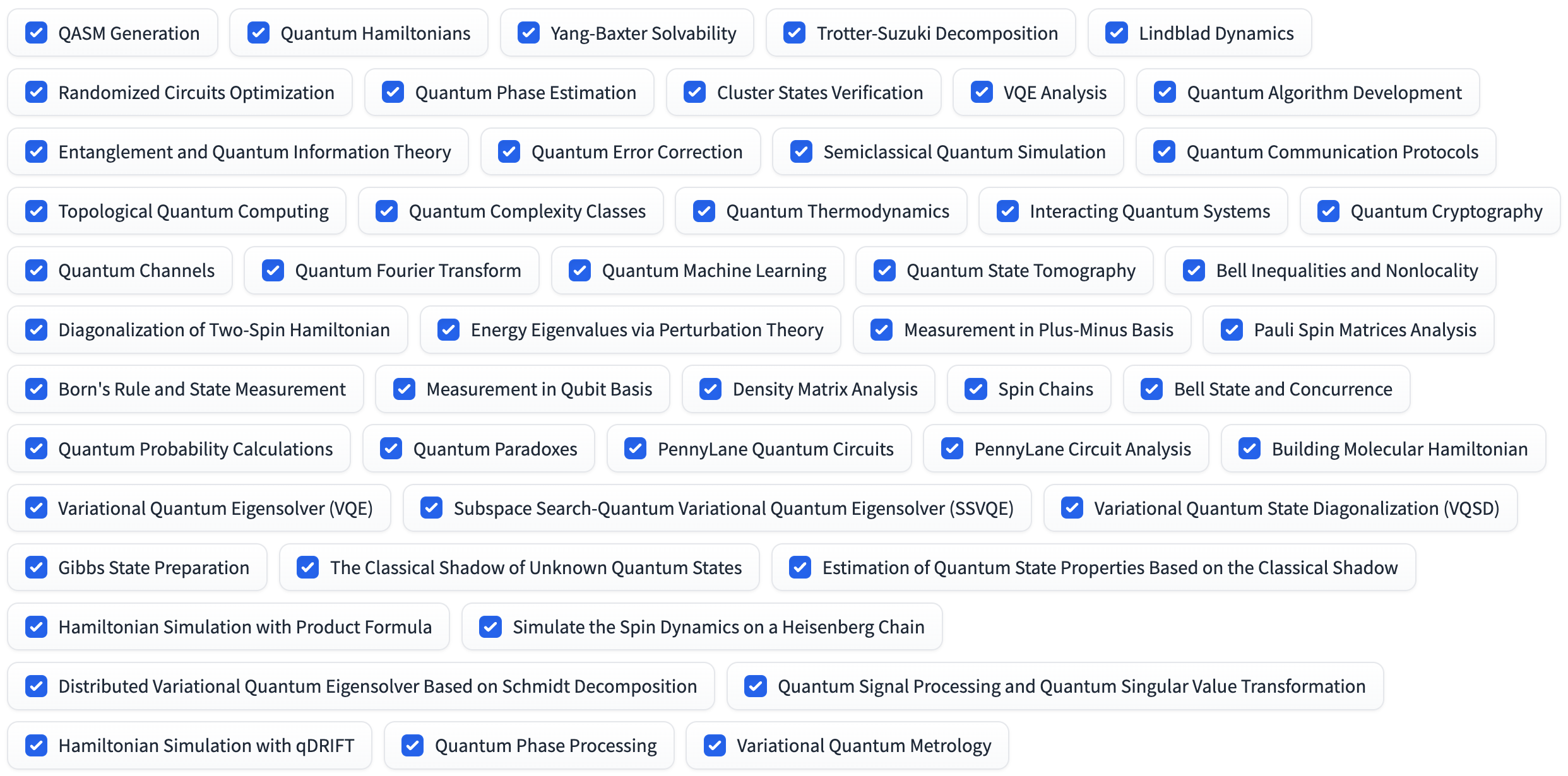}
	\caption{Partial list of the core domains of QuantumLLMInstruct. QLMMI leverages predefined templates to ensure domain relevance and mathematical precision in critical areas such as quantum phase estimation, Hamiltonian analysis, and randomized circuit optimization. Beyond these predefined categories, LLMs autonomously generated hundreds of novel quantum computing domains, highlighting their adaptability and scalability.}
	\label{fig:main-domains}
\end{figure}

To address these challenges, we introduce \textbf{QuantumLLMInstruct (QLMMI)}, a comprehensive dataset and framework (Fig.~\ref{fig:db-flow}) leveraging state-of-the-art LLMs for quantum problem-solving. QLMMI automates the generation of quantum problem statements, derives solutions, and refines outputs through advanced reasoning techniques and self-assessment mechanisms~\cite{madaan2024self, wei2022chain, zhou2024self}. By doing so, QLMMI establishes a robust foundational resource for tackling a wide range of quantum computing tasks (Fig.~\ref{fig:main-domains}) with minimal human intervention.

\begin{figure}[h]
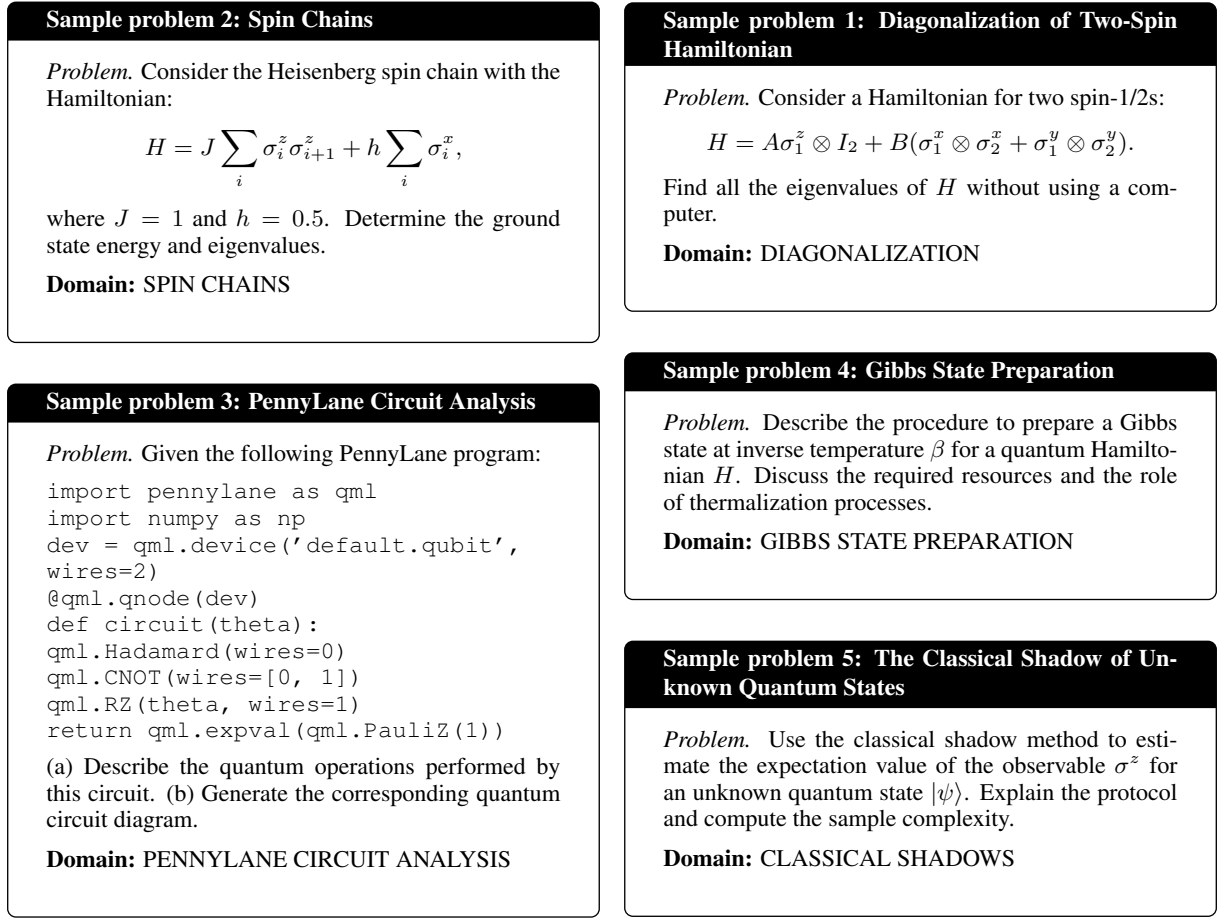

	\centering
	\begin{minipage}{0.49\textwidth}
		\begin{tcolorbox}[
			title=\textbf{\small Sample problem 2: Spin Chains},
			colframe=black,
			boxrule=0.5pt,
			colback=white,
			equal height group=probgroup1
			]
			\footnotesize
			\textit{Problem.} Consider the Heisenberg spin chain with the Hamiltonian:
			\[
			H = J \sum_{i} \sigma_i^z \sigma_{i+1}^z + h \sum_i \sigma_i^x,
			\]
			where \( J = 1 \) and \( h = 0.5 \). Determine the ground state energy and eigenvalues.\par
			\vspace{0.5\baselineskip}
			\textbf{Domain:} SPIN CHAINS\par
			\vspace{\baselineskip}
		\end{tcolorbox}

		\vspace{0.15cm}

		\begin{tcolorbox}[
			title=\textbf{\small Sample problem 3: PennyLane Circuit Analysis},
			colframe=black,
			boxrule=0.5pt,
			colback=white
			]
			\footnotesize
			\textit{Problem.} Given the following PennyLane program:
			\begin{verbatim}
				import pennylane as qml
				import numpy as np
				dev = qml.device('default.qubit',
				wires=2)
				@qml.qnode(dev)
				def circuit(theta):
				qml.Hadamard(wires=0)
				qml.CNOT(wires=[0, 1])
				qml.RZ(theta, wires=1)
				return qml.expval(qml.PauliZ(1))
			\end{verbatim}
			(a) Describe the quantum operations performed by this circuit. (b) Generate the corresponding quantum circuit diagram.\par
			\vspace{0.5\baselineskip}
			\textbf{Domain:} PENNYLANE CIRCUIT ANALYSIS\par
			\vspace{\baselineskip}
		\end{tcolorbox}
	\end{minipage}%
	\hfill%
	\begin{minipage}{0.49\textwidth}
		\begin{tcolorbox}[
			title=\textbf{\footnotesize Sample problem 1: Diagonalization of Two-Spin Hamiltonian},
			colframe=black,
			boxrule=0.5pt,
			colback=white
			]
			\footnotesize
			\textit{Problem.} Consider a Hamiltonian for two spin-1/2s:
			\[
			H = A \sigma_1^z \otimes I_2 + B (\sigma_1^x \otimes \sigma_2^x + \sigma_1^y \otimes \sigma_2^y).
			\]
			Find all the eigenvalues of \( H \) without using a computer.\par
			\vspace{0.5\baselineskip}
			\textbf{Domain:} DIAGONALIZATION\par
			\vspace{\baselineskip}
		\end{tcolorbox}

		\vspace{0.15cm}

		\begin{tcolorbox}[
			title=\textbf{\small Sample problem 4: Gibbs State Preparation},
			colframe=black,
			boxrule=0.5pt,
			colback=white
			]
			\footnotesize
			\textit{Problem.} Describe the procedure to prepare a Gibbs state at inverse temperature \( \beta \) for a quantum Hamiltonian \( H \). Discuss the required resources and the role of thermalization processes.\par
			\vspace{0.5\baselineskip}
			\textbf{Domain:} GIBBS STATE PREPARATION\par
			\vspace{\baselineskip}
		\end{tcolorbox}

		\vspace{0.15cm}

		\begin{tcolorbox}[
			title=\textbf{\footnotesize Sample problem 5: The Classical Shadow of Unknown Quantum States},
			colframe=black,
			boxrule=0.5pt,
			colback=white
			]
			\footnotesize
			\textit{Problem.} Use the classical shadow method to estimate the expectation value of the observable \( \sigma^z \) for an unknown quantum state \( |\psi\rangle \). Explain the protocol and compute the sample complexity.\par
			\vspace{0.5\baselineskip}
			\textbf{Domain:} CLASSICAL SHADOWS\par
			\vspace{\baselineskip}
		\end{tcolorbox}
	\end{minipage}
	\caption{\textbf{QuantumLLMInstruct} encompasses a wide range of quantum physics problems meticulously designed using carefully crafted templates (Fig.~\ref{fig:main-domains}), ensuring domain specificity and mathematical rigor. Spanning over \textbf{90 predefined quantum computing domains} (see Sec.~\ref{sec:domains})—including quantum cryptography, spin chain models, and Trotter-Suzuki decompositions~\cite{Ising1925, Heisenberg1928, Suzuki1976} -- the dataset guarantees both diversity and precision. Beyond these templates, the LLMs demonstrated emergent capabilities, autonomously generating hundreds of new quantum domains without explicit prompt engineering~\cite{wei2022chain, wei2022emergent, madaan2024self}, establishing QLMMI as a versatile and scalable resource in quantum problem-solving. These representative samples illustrate the dataset's depth in problem variety, rigor, and domain-specific relevance.}
	\label{fig:quantum-llm-problems}
\end{figure}

The foundation of QLMMI lies in its reliance on carefully crafted templates (Fig.~\ref{fig:main-domains}), ensuring domain specificity and mathematical rigor. These templates span over \textbf{90 predefined quantum computing domains} (see Sec.~\ref{sec:domains}), including quantum cryptography, spin chain models, and Trotter-Suzuki decompositions~\cite{Ising1925, Heisenberg1928, Suzuki1976}. While ensuring both diversity and precision, the dataset extends far beyond these predefined categories. Remarkably, LLMs demonstrated emergent capabilities, \textbf{autonomously generating hundreds of quantum sub-domains without explicit prompt engineering}~\cite{wei2022chain, wei2022emergent, madaan2024self}, underscoring the scalability and adaptability of LLMs in quantum problem-solving. This positions QLMMI as a versatile and comprehensive resource.

The dataset employs detailed natural language prompts paired with LaTeX-formatted mathematical expressions, ensuring clarity and precision. Insights from existing resources like Omni-MATH~\cite{omnimath2024}, OpenMathInstruct-1~\cite{openmath2024}, and MetaMath~\cite{yu2023metamath} were integrated to address challenges in quantum reasoning, circuit optimization, and algorithm design. Moreover, open-source LLMs like Qwen-2.5-Coder~\cite{yang2024qwen2} were utilized to circumvent restrictive licensing, ensuring robust reasoning capabilities while promoting accessibility and scalability.

Here, (Fig.~\ref{fig:quantum-llm-problems}) we present a representative list of quantum computing problem types from the QLMMI dataset, highlighting its depth, rigor, and domain-specific relevance.

\subsection{Challenges of Synthetic Data Generation vs. Training and Fine-Tuning}

This work \textbf{focuses exclusively on the generation of synthetic data for quantum computing tasks}, deliberately excluding training or fine-tuning of large LLMs. No trained model is provided in this study. Instead, we prioritize creating a high-quality dataset to support future research without the computational demands of training or fine-tuning.

Fine-tuning large LLMs, such as GPT-4, requires extensive engineering and computational infrastructure~\cite{achiam2023gpt}. Studies~\cite{wei2022chain, wang2022selfconsistency} further emphasize that the success of fine-tuning is contingent on precise data preparation and hyperparameter configuration. By concentrating on synthetic data generation, this work aligns with broader initiatives to democratize AI access through the use of pre-trained models and high-quality datasets~\cite{yu2023metamath, hendrycks2021measuring}. Below, we outline the primary challenges that influenced this decision:

\paragraph{Resource Requirements.}

Fine-tuning LLMs like Qwen2.5-Math-72B necessitates high-performance GPU clusters (e.g., 8 x H100), which are prohibitively expensive and inaccessible to many researchers and smaller institutions~\cite{yang2024qwen2, achiam2023gpt}. The iterative nature of fine-tuning compounds these costs, requiring repeated runs to optimize model performance~\cite{wei2022chain, yao2024tree}. Additionally, generating high-quality fine-tuning datasets demands domain expertise to avoid introducing errors or biases~\cite{touvron2023llama, du2022glm, goodfellow2016deep}. Evaluation complexities further complicate the process, especially for niche domains like quantum computing, where datasets such as QDataset~\cite{qdataset2021} or MNISQ~\cite{mnisq2023} are limited in scope and availability~\cite{gao2023framework}.

\paragraph{Expertise Barriers.}

Developing fine-tuning datasets for specialized fields like quantum physics requires significant domain expertise~\cite{touvron2023llama, brown2020language}. Errors in dataset preparation or training methodologies can degrade model performance, while robust evaluation frameworks are often overlooked~\cite{wei2022emergent, wang2022selfconsistency}, leading to suboptimal outcomes.

\paragraph{Evaluation Complexities.}

Evaluation of fine-tuned models is particularly challenging in niche domains like quantum computing. Models frequently exhibit inconsistent performance across datasets or evaluation contexts~\cite{gao2023framework}, and the scarcity of task-specific datasets, such as QDataset~\cite{qdataset2021} or MNISQ~\cite{mnisq2023}, exacerbates these challenges.

\paragraph{Why Focus Solely on Synthetic Data Generation?}

Considering these challenges, this work emphasizes synthetic data generation as a cost-efficient and reproducible alternative. Generating synthetic data is significantly less resource-intensive than fine-tuning, enabling scalable dataset creation~\cite{yue2023mammoth}. Moreover, synthetic data pipelines are transparent, hardware-independent, and consistent, ensuring reproducibility~\cite{achiam2023gpt}. Additionally, high-quality datasets empower researchers with limited resources to explore quantum AI, fostering broader accessibility and democratization~\cite{toshniwal2024openmathinstruct}. By focusing on synthetic data generation, this work establishes a robust foundation for quantum computing research, enabling the exploration of advanced tasks with pre-trained models while circumventing the computational and expertise barriers associated with fine-tuning.

\section{Core Features of QuantumLLMInstruct (QLMMI)}

The defining attributes of QLMMI include its extensive domain coverage and a meticulously designed four-stage creation process, ensuring both breadth and depth in quantum problem-solving:

\begin{itemize}
	\item \textbf{Stage I: Foundational Problem Generation.} The cornerstone of QLMMI lies in the use of meticulously crafted templates (see Fig.~\ref{fig:main-domains}) to ensure domain specificity and mathematical precision. These templates facilitate the generation of problems across diverse quantum computing domains, encompassing tasks such as synthetic Hamiltonians, QASM code generation, Jordan-Wigner transformations, Trotter-Suzuki decompositions, quantum phase estimation, variational quantum eigensolvers (VQE), cluster state verification, randomized circuit optimization, spin dynamics on Heisenberg chains, Yang-Baxter solvability, quantum signal processing, and Gibbs state preparation for quantum thermodynamics. Classical models like Ising~\cite{Ising1925} and Heisenberg spin chains~\cite{Heisenberg1928} serve as central elements in these tasks, ensuring mathematical rigor and applicability. Notably, QLMMI extends far beyond these predefined templates, as LLMs exhibit emergent capabilities to autonomously generate hundreds of new quantum domains without explicit prompt engineering. This adaptability highlights the scalability of LLMs in quantum problem-solving, positioning QLMMI as a highly versatile and comprehensive resource. For this stage, the \textbf{Qwen-2-coder-instruct} model is employed, offering exceptional precision in generating domain-specific problems.

	\item \textbf{Stage II: Solution Generation.} To ensure the dataset's completeness and accuracy, solutions are developed for the generated problems. This stage leverages the \textbf{Qwen-2-coder-instruct} model, which is optimized for crafting detailed and contextually accurate solutions tailored to quantum computing challenges.

	\item \textbf{Stage III: Dataset Enrichment.} Advanced reasoning frameworks, such as Chain-of-Thought (CoT)~\cite{wei2022chain} and Task-Oriented Reasoning and Action (ToRA)~\cite{wu2023tora}, are integrated to enhance the quality and diversity of problem-solution pairs. These methods ensure that the dataset covers a wide spectrum of quantum challenges, improving both its depth and adaptability. The \textbf{Qwen-2-math-instruct} model, renowned for its mathematical precision and reasoning prowess, is utilized in this stage to enrich the dataset effectively.

	\item \textbf{Stage IV: Validation and Quality Assurance.} The final stage applies robust validation through a Judge LLM employing self-critique techniques~\cite{madaan2024self} to ensure the reliability and accuracy of the dataset. This process significantly minimizes human oversight while maintaining high-quality standards. The \textbf{Qwen-2-math-instruct} model, with its expertise in evaluating mathematical correctness and logical coherence, is employed to rigorously assess the entries, guaranteeing a dependable and accurate dataset.
\end{itemize}

\paragraph{Contributions}

This work introduces several key contributions:
\begin{enumerate}
	\item \textbf{Comprehensive Dataset:} A scalable dataset containing over 500,000 rigorously curated quantum problem-solution pairs designed for fine-tuning LLMs for quantum-specific tasks. An example question is depicted in (Fig.~\ref{fig::example-instruction}).
	\item \textbf{Four-Stage Workflow:} A robust framework combining predefined templates, Chain-of-Thought (CoT) reasoning~\cite{wei2022chain}, Task-Oriented Reasoning and Action (ToRA)~\cite{wu2023tora}, and self-assessment mechanisms~\cite{madaan2024self} for automated quality assurance.
	\item \textbf{Scalable Infrastructure:} Integration of tools like DuckDB and Gradio enables efficient dataset management and interactive exploration of quantum problems~\cite{bergholm2018pennylane}.
	\item \textbf{Open Access:} QLMMI provides public access to raw datasets, data generation code, and training scripts, fostering collaboration in quantum computing research.
\end{enumerate}

\paragraph{Future Directions}

Future work aims to extend QLMMI's impact through:
\begin{itemize}
	\item \textbf{Advanced Model Fine-Tuning:} Expanding evaluations on models such as Qwen-2.5-Math and Qwen-2.5-Code-Instruct to further enhance quantum reasoning capabilities~\cite{yang2024qwen2, luo2023wizardmath}.
	\item \textbf{Integration with Retrieval-Augmented Models:} Incorporating advanced retrieval mechanisms to improve contextual understanding for complex quantum tasks~\cite{gao2023framework, shi2023replug}.
	\item \textbf{Cross-Domain Applications:} Extending problem templates and solutions to intersect with fields like quantum chemistry and cryptography, broadening QLMMI’s interdisciplinary impact~\cite{Preskill2018, Lloyd1996}.
\end{itemize}

\section{Related Work}

The development of large-scale datasets for mathematical and scientific reasoning has been a key focus in recent years. In (Tab.~\ref{tab:combined-datasets-comparison}) we highlight some of the recent works synthesizing datasets in mathematics and physics.

\begin{table}[h!]
	\centering
	\footnotesize
	\begin{tabular}{lccc}
		\toprule
		\textbf{Dataset} & \textbf{Focus/Domain} & \textbf{Size/Scale} & \textbf{Methodology} \\
		\midrule
		MathInstruct \cite{yue2023mammoth} & Math/Instruction Tuning & 1.8M examples & CoT and PoT \\
		MetaMathQA \cite{yu2023metamath} & Enhanced Reasoning/Math & 100K examples & Fine-tuning with GSM8K \\
		OpenMathInstruct \cite{toshniwal2024openmathinstruct} & Synthetic Math Solutions/Math & 1.8M examples & Hybrid instruction tuning \\
		WizardCoder \cite{luo2023wizardcoder} & Code-Specific Reasoning/Math & 200K+ examples & Evol-instruct with StarCoder \\
		Lila \cite{lila2023} & Math Benchmark & 272K examples & - \\
		MathCodeInstruct \cite{mathcodeinstruct2023} & Math with Code & 80K examples & GPT-4 + Self \\
		ToRA \cite{tora2023} & Task-Oriented Math Reasoning & 16K examples & GPT-4 \\
		OMNI-MATH \cite{omnimath2024} & Olympiad-Level Math & N/A & GPT-3.5 \\
		QDataset \cite{qdataset2021} & Quantum Computing/Physics & 52 datasets & - \\
		MNISQ \cite{mnisq2023} & Quantum Numerical Simulations & 4.95M datapoints & - \\
		QCircuitNet \cite{qcircuitnet2024} & Quantum Algorithms/Physics & N/A & Closed-source LLMs \\
		\rowcolor[gray]{0.9} \textbf{QuantumLLMInstruct (ours)} & \textbf{Broad Quantum Domains} & \textbf{500K examples} & \textbf{Qwen-Based} \\
		\bottomrule
	\end{tabular}
	\caption{Comparison of datasets for quantum and mathematical reasoning, highlighting size, generating methodology, and focus domain. QLMMI sets itself apart with its extensive domain coverage and quantum-specific tasks.}
	\label{tab:combined-datasets-comparison}
\end{table}

\paragraph{Mathematical and Physics Datasets.}

In recent years, there has been growing interest in leveraging large language models (LLMs) to address tasks in quantum computing, mathematical reasoning, and physics. These efforts have resulted in diverse contributions, as detailed below.

\paragraph{Instruction Tuning.}

Instruction tuning has emerged as a pivotal technique for tailoring LLMs to specific tasks. Two primary approaches to instruction tuning have been explored:
\begin{itemize}[leftmargin=*]
	\item \textit{Human-Written Data:} Frameworks such as FLAN~\cite{wei2021finetuned}, T0~\cite{sanh2021multitask}, and SuperNI~\cite{wang2022super} rely on pre-existing human-labeled datasets. While these datasets offer high reliability, their size and diversity are constrained by the high cost of manual annotation.
	\item \textit{Synthesized Data:} Models like Self-Instruct~\cite{wang2023self}, WizardLM~\cite{xu2023wizardlm}, and GPT4-Alpaca~\cite{peng2023instruction} generate datasets via powerful LLMs like GPT-4~\cite{achiam2023gpt}. This approach scales easily but may introduce hallucinations, reducing data quality. The diversity of synthesized data also depends on the comprehensiveness of the seed dataset.
\end{itemize}

QLMMI adopts a synthesized data approach, leveraging instruction-tuned LLMs to generate, refine, and validate quantum problems. Advanced filtering mechanisms mitigate issues like noise and hallucination, ensuring high-quality instruction-tuning pairs. This aligns with efforts like OpenMathInstruct~\cite{toshniwal2024openmathinstruct} and MetaMATH~\cite{yu2023metamath}, which demonstrate the efficacy of synthesized datasets in improving LLM reasoning capabilities.

\paragraph{Mathematical Reasoning.}

Mathematical reasoning has seen substantial advancements through innovative prompting strategies, continued training, and instruction tuning.

\begin{figure}[h!]
	\centering
	\includegraphics[width=\textwidth]{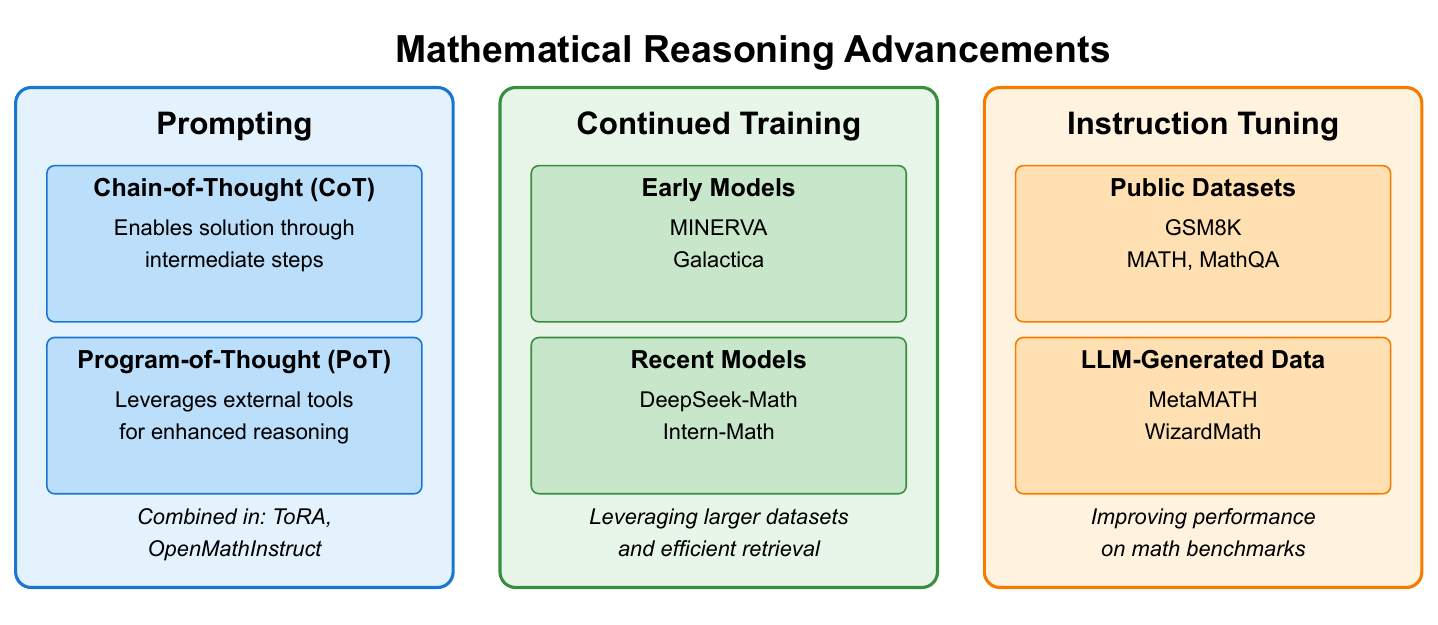}
	\caption{Overview of mathematical reasoning advancements in LLMs. \textbf{Prompting} techniques include Chain-of-Thought (CoT)~\cite{nye2022show,wei2022chain} for intermediate reasoning steps and Program-of-Thought (PoT)~\cite{chen2023program,gao2023pal} for tool-augmented reasoning, combined in models like ToRA~\cite{gou2023tora} and OpenMathInstruct~\cite{toshniwal2024openmathinstruct}. \textbf{Continued Training} progressed from early models like MINERVA~\cite{lewkowycz2022solving} and Galactica~\cite{taylor2022galactica} to advanced models such as DeepSeek-Math~\cite{shao2024deepseekmath} and Intern-Math~\cite{ying2024internlm}. \textbf{Instruction Tuning} leverages both public datasets (e.g., GSM8K~\cite{cobbe2021training}, MATH~\cite{hendrycks2021measuring}, MathQA~\cite{amini2019mathqa}) and LLM-generated data (MetaMATH~\cite{yu2023metamath}, WizardMath~\cite{luo2023wizardmath}) to enhance mathematical capabilities.}
	\label{fig:math-advances}
\end{figure}

QLMMI builds on these strategies by integrating CoT reasoning and PoT methodologies into its dataset generation pipeline, enhancing LLMs’ ability to tackle mathematical challenges across quantum computing domains.

\paragraph{Physics Reasoning.}

Physics reasoning remains underexplored compared to mathematics. Benchmarks like MMLU~\cite{hendrycks2020measuring}, SciEval~\cite{sun2023scieval}, and GPQA~\cite{rein2023gpqa} provide evaluation frameworks but lack large-scale datasets tailored to physics.

Techniques like CoT~\cite{wei2022cot} and PoT~\cite{chen2022pot} have shown promise in physics, particularly when integrated with intermediate validation~\cite{li2022stepaware} and external computational tools. Fine-tuned models like WizardCoder~\cite{luo2023wizardcoder} demonstrate the potential of combining natural language reasoning with code generation, emphasizing the importance of domain-specific tuning.

A notable contribution in quantum computing is QCircuitNet~\cite{qcircuitnet2024}, employing closed-source LLMs for a comprehensive range of quantum algorithms, from foundational primitives to advanced applications. However, QCircuitNet focuses on algorithm design, overlooking the broader spectrum of quantum domains addressed by QLMMI. Similarly, ManyBodyLLMs~\cite{manybodyllms2024} explores GPT-4's ability to perform graduate-level quantum many-body physics calculations, achieving 87.5\% adherence across Hartree-Fock mean-field theory tasks. While depth-focused, these works lack the breadth offered by QLMMI.

PhysQA~\cite{ding2023physqa} emphasizes elementary physics education through GPT-3.5-generated solutions to high school physics problems. It contrasts with QLMMI’s focus on advanced quantum domains.

In contrast to these efforts, QLMMI integrates advanced reasoning techniques like CoT~\cite{wei2022chain, wei2022cot} and self-critique~\cite{madaan2024self, saunders2022self} to ensure robust problem-solving and high-quality validation. It spans domains like synthetic Hamiltonians and Trotter-Suzuki decompositions~\cite{Suzuki1976}, making it a versatile resource for addressing challenges in quantum computing.

\begin{enumerate}
	\item \textbf{Instruction Generation:} One aligned LLM generates LaTeX-formatted quantum physics problems adhering to strict guidelines.
	\item \textbf{Solution Generation:} A second LLM generates detailed solutions for these problems, enabling specialized fine-tuning of both problem and solution generation.
\end{enumerate}

For a full list of instruction-tuned datasets, refer to Appendix~\ref{apnd:add-dbb}.

\section{Data Generation Pipeline}

The prompts used at various stages are meticulously crafted to align with the capabilities of LLMs while ensuring consistency in the format of problems and solutions.

\begin{figure}[h!]
	\centering
	\includegraphics[width=\textwidth]{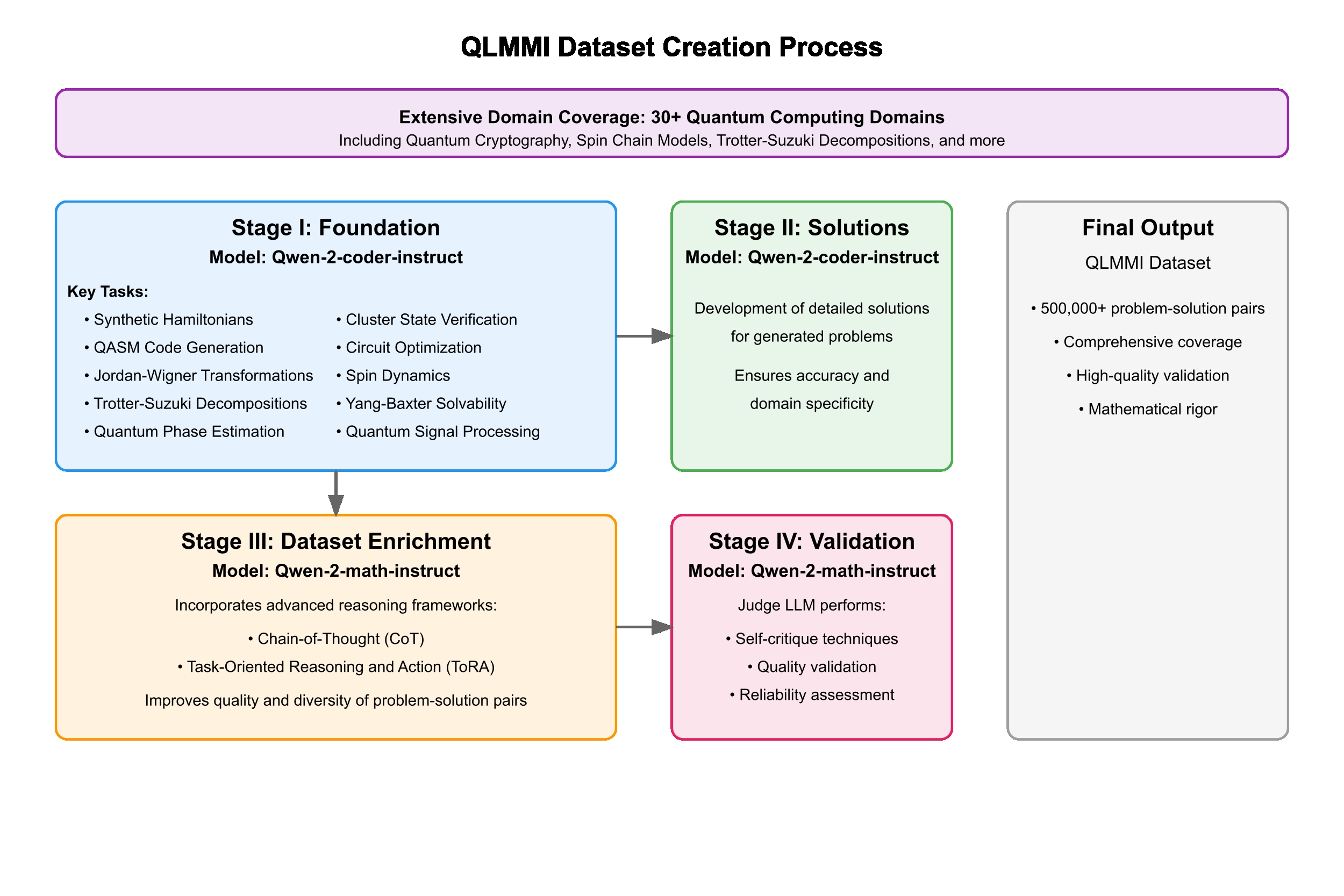}
	\caption{Dataset Creation Workflow for QuantumLLMInstruct. The process comprises four stages: \textbf{Stage I} generates foundational problems with LLMs, focusing on tasks such as synthetic Hamiltonians, QASM code, Jordan-Wigner transformations, and Trotter-Suzuki decompositions for models like Ising~\cite{Ising1925} and Heisenberg spin chains~\cite{Heisenberg1928}. \textbf{Stage II} develops solutions to these problems using LLMs. \textbf{Stage III} enriches the dataset using Chain-of-Thought (CoT)~\cite{wei2022chain} and Task-Oriented Reasoning and Action (ToRA)~\cite{wu2023tora}. \textbf{Stage IV} validates the dataset through self-critique mechanisms~\cite{madaan2024self}, performed by a Judge LLM, ensuring high-quality entries.}
	\label{fig:db-flow-detail}
\end{figure}

\subsection{LLM Models}
\label{sec:llm_models}

The Qwen family of Large Language Models (LLMs) was employed for both instruction and response generation. These models are tailored to handle complex quantum physics tasks, providing high precision and adaptability across diverse quantum domains. The Qwen family offers models ranging from lightweight variants to high-performance versions, accommodating a broad spectrum of computational needs.

\begin{table}[h!]
	\centering
	\footnotesize
	\begin{tabular}{lccc}
		\toprule
		\textbf{Model Name} & \textbf{Parameters (Billion)} & \textbf{Dataset Trained On} & \textbf{Focus Domain} \\
		\midrule
		Coder-1.5B-Instruct \cite{qwen2024} & 1.5 & Code, Mathematical Problems & Instruction Tuning \\
		Coder-3B-Instruct \cite{qwen2024} & 3 & Code, Scientific Papers & Instruction Tuning \\
		Coder-7B-Instruct \cite{qwen2024} & 7 & Code and Mathematical Problems & Advanced Coding Tasks \\
		Math-7B-Instruct \cite{yang2024qwen2} & 7 & Mathematics-Focused Datasets & Math Problem Solving \\
		Coder-32B-Instruct \cite{qwen2024} & 32 & Extensive Code and Scientific Datasets & Complex Scientific Tasks \\
		Llama-3.2B-Instruct \cite{touvron2023llama} & 3 & General-Purpose Instruction Dataset & General Reasoning \\
		\bottomrule
	\end{tabular}
	\caption{Details of LLMs used in the Quantum Physics Problem Generator. Each model is characterized by its parameter count, training datasets, and target focus domain.}
	\label{tab:llm_models_details}
\end{table}

\paragraph{Flexibility and Scalability}

The Qwen2.5 models (Table~\ref{tab:llm_models_details}) provide parameter configurations ranging from 0.5B to 72B, enabling users to tailor model usage based on available GPU memory and specific requirements. This ensures scalability for both research and production applications.

\paragraph{Usage in Instruction and Response Generation}

\begin{itemize}
	\item \textbf{Instruction Generation:} Qwen2.5-Coder-*B-Instruct models were used to create problem prompts, benefiting from their expertise in coding and mathematical formulation.
	\item \textbf{Response Generation:} Qwen2.5-Math-*B-Instruct models generated step-by-step solutions with LaTeX-formatted precision, ensuring clarity and rigor.
\end{itemize}

\subsection{Stage 1: Problem Generation}

In this stage, the first LLM, a  Qwen2.5-Coder-*B-Instruct, is prompted to generate a quantum physics problem in LaTeX format. Each problem adheres to predefined domain types (see Sec.~\ref{sec:domains}), ensuring solvability and the inclusion of necessary mathematical precision. The following structured prompt is employed:

\begin{figure}[h!]
	\definecolor{brightgray}{HTML}{A9A9A9}
	\begin{tcolorbox}[colback=brightgray!5!white,colframe=brightgray!20!black]
		\begin{footnotesize}
			\textbf{Illustrative Example: Generating a Quantum Physics Problem}
			\Sepline
			\begin{minipage}[t]{0.32\linewidth}
				\textbf{Instruction Prompt} \\
				Generate a single detailed quantum physics problem for an exam in LaTeX format. Do not solve the problem. \\
				\begin{itemize}
					\item Avoid additional explanations or comments outside of LaTeX.
					\item Exclude unnecessary LaTeX imports (e.g., \texttt{\textbackslash documentclass\{\}}, \texttt{\textbackslash usepackage\{\}}, or \texttt{\textbackslash begin\{document\}}).
					\item Format all mathematical equations and symbols strictly in LaTeX.
				\end{itemize}
			\end{minipage}
			\begin{minipage}[t]{0.58\linewidth}
				\textbf{Output Format and Example} \\
				\begin{enumerate}
					\item \textbf{Problem:} Clearly define the quantum physics problem using LaTeX for mathematical precision. Provide all equations or descriptions necessary for solving the problem. For instance:
					\[
					H = - \sigma_1^z \sigma_2^z - 0.5 (\sigma_1^x + \sigma_2^x)
					\]
					\item \textbf{Domain:} Include a concise two-word domain description in CAPS such as \texttt{ISING HAMILTONIAN}.
					\item Ensure the problem strictly adheres to one of the following predefined domain types:
					\begin{itemize}
						\item Synthetic Hamiltonians
						\item Quantum Spin Chains
						\item Yang-Baxter Solvability
						\item Quantum Circuit Design
					\end{itemize}
				\end{enumerate}
			\end{minipage}
		\end{footnotesize}
	\end{tcolorbox}
	\caption{\small Step-by-step instruction for generating a detailed quantum physics problem in LaTeX format. The framework dynamically selects from over 90 quantum physics domains (see Sec.~\ref{sec:domains}), leveraging LLMs to autonomously generate sub-domains, thus expanding to thousands of unique quantum problem categories.}
	\label{fig::example-instruction}
\end{figure}

\subsection{Stage 2: Solution Generation}

The output of Stage 1 is provided to a second Qwen2.5-Math-*B-Instruct LLM tasked with generating step-by-step solutions such as the one depicted in Figure \ref{tab:hamiltonians_analytical2}. The following prompt (Fig. \ref{fig::solution-prompt}) guides this stage:

\begin{figure}[h!]
	\definecolor{brightgray}{HTML}{A9A9A9}
	\begin{tcolorbox}[colback=brightgray!5!white,colframe=brightgray!35!black]
		\begin{footnotesize}
			\textbf{Illustrative Example: Solution Prompt for Quantum Physics Problems}
			\Sepline
			\begin{minipage}[t]{0.32\linewidth}
				\textbf{Solution Instruction Prompt} \\
				You are an expert in quantum physics and must provide detailed solutions to quantum physics problems. \\
				\begin{itemize}
					\item All solutions must be in plain text format.
					\item Use LaTeX to represent all mathematical equations and symbols.
					\item Ensure clarity, accuracy, and completeness in each step of the solution.
				\end{itemize}
			\end{minipage}
			\begin{minipage}[t]{0.58\linewidth}
				\textbf{Output Format and Example} \\
				\begin{enumerate}
					\item \textbf{Solution Prompt:} Provide a complete solution to the following quantum physics problem in plain text format:
					\[
					{problem}
					\]
					\item \textbf{Expected Response:}
					\begin{itemize}
						\item Step-by-step derivation with appropriate mathematical rigor.
						\item Use LaTeX for all equations and symbols, for example:
						\[
						H = \begin{bmatrix}
							1 & 0 \\\\
							0 & -1
						\end{bmatrix}.
						\]
						\item A final summary or conclusion, if applicable.
					\end{itemize}
				\end{enumerate}
			\end{minipage}
		\end{footnotesize}
	\end{tcolorbox}
	\caption{\small Solution \textbf{prompt} for quantum physics problems, requiring step-by-step solutions with all mathematical symbols and equations in LaTeX format. Emphasizes clarity, accuracy, and completeness in each solution.}
	\label{fig::solution-prompt}
\end{figure}

\paragraph{Example: Ising Model Hamiltonian}

Here (Fig.~\ref{tab:hamiltonians_analytical2}) we present a practical example from the dataset, illustrating a simplified Ising model for two qubits with interaction strength \( J = 1 \) and transverse field \( h = 0.5 \). This solution was generated as a response to the solution-prompt depicted in (Fig.~\ref{fig::solution-prompt}).

\begin{figure}[h!]
	\definecolor{brightgray}{HTML}{A9A9A9}
	\begin{tcolorbox}[colback=brightgray!5!white,colframe=brightgray!35!black]
		\begin{footnotesize}
			\textbf{Illustrative Example: Solving a Sample Hamiltonian}
			\Sepline
			\begin{minipage}[t]{0.32\linewidth}
				\textbf{Problem} \\
				Analytically solve the Hamiltonian \\
				\[
				H = - \sigma_1^z \sigma_2^z - 0.5 (\sigma_1^x + \sigma_2^x)
				\]
			\end{minipage}
			\begin{minipage}[t]{0.58\linewidth}
				\textbf{Step-by-Step Solution} \\
				1. Express the Hamiltonian in matrix form:
				\[
				H = \begin{pmatrix}
					-1 & -0.5 & -0.5 & 0 \\
					-0.5 & 1 & 0 & -0.5 \\
					-0.5 & 0 & 1 & -0.5 \\
					0 & -0.5 & -0.5 & -1
				\end{pmatrix}
				\]
				2. Set up the eigenvalue equation \( \det(H - \lambda I) = 0 \):
				\[
				\det\begin{pmatrix}
					-1 - \lambda & -0.5 & -0.5 & 0 \\
					-0.5 & 1 - \lambda & 0 & -0.5 \\
					-0.5 & 0 & 1 - \lambda & -0.5 \\
					0 & -0.5 & -0.5 & -1 - \lambda
				\end{pmatrix} = 0
				\]
				3. Solve the characteristic equation:
				\[
				\lambda^4 + \lambda^3 - 3\lambda^2 - \lambda + 1 = 0
				\]
				4. Eigenvalues (approximate):
				\[
				\lambda_1 = -1.618, \quad \lambda_2 = 0, \quad \lambda_3 = 0, \quad \lambda_4 = 1.618
				\]
			\end{minipage}
		\end{footnotesize}
	\end{tcolorbox}
	\caption{\small Step-by-step \textbf{solution} for solving the Hamiltonian \( H = - \sigma_1^z \sigma_2^z - 0.5 (\sigma_1^x + \sigma_2^x) \) analytically, including eigenvalues and matrix representation.}
	\label{tab:hamiltonians_analytical2}
\end{figure}

\paragraph{Stage 3: Problem-Solution Pair Dataset Enrichment}

In this stage, we enrich the dataset by leveraging advanced reasoning methodologies (Fig.~\ref{fig:stage3-cot}), (Fig.~\ref{fig:stage3-cop}), such as CoT  \cite{wei2022chain}, ToRA, and CoP \cite{wei2022chain}. These techniques, traditionally applied in fields like combinatorics and optimization, are extended into quantum computing to address tasks such as Hamiltonian decomposition, QASM code generation, and eigenvalue computations. Existing datasets, including CoT and ToRA, offer problems with clear, unambiguous solutions, which serve as templates for analogous quantum computing challenges.

\begin{figure}[h!]
	\centering
	\begin{minipage}{\textwidth}
		\begin{tcolorbox}[
			title=\textbf{\footnotesize Sample Prompt: Physics Problem-Solution Verification Using ToRA and CoT Frameworks},
			colframe=black,
			boxrule=0.5pt,
			colback=white
			]
			\footnotesize
			\textit{Prompt.} You are a quantum physics solving assistant. You will be presented with a physics problem and a proposed solution. Your task is to determine if the provided solution is correct. If the solution is correct, confirm it with a detailed explanation. If the solution is incorrect, identify the errors and provide a full corrected solution using the ToRA framework: Think, understand and restate the problem, identifying key details; Organize, break it into manageable steps; Reason, solve systematically, showing all intermediate steps and calculations; and Answer, present the final solution clearly and concisely. Always use Chain of Thought reasoning to ensure logical progression and clarity.

			\textbf{Example:}

			\textit{Problem:} A train travels 120 miles in 2 hours. What is its average speed?

			\textit{Proposed Solution:} The average speed is 70 mph.

			\textbf{Analysis:} The solution is incorrect. Think: Speed is distance divided by time. The distance is 120 miles, and the time is 2 hours. Organize: Write the formula Speed = Distance / Time. Reason: Substitute the values, Speed = 120 / 2 = 60 mph. Answer: The correct average speed is 60 mph.

			\textbf{Now analyze the following problem and solution:}

			\textit{Problem: [Insert Problem Here]} \\
			\textit{Proposed Solution: [Insert Proposed Solution Here]} \\
		\end{tcolorbox}
	\end{minipage}
	\caption{\textbf{Stage 3} Sample prompt - Enrichment using ToRA and CoT frameworks for quantum physics problem verification.}
	\label{fig:stage3-cot}
\end{figure}

The enriched dataset bridges the gap between raw task definitions and structured problem-solution pairs. By utilizing techniques such as CoT, ToRA, and CoP, we enhance both the diversity and applicability of the dataset, ensuring its relevance for advanced quantum computing tasks.

\begin{figure}[h!]
	\centering
	\begin{minipage}{\textwidth}
		\begin{tcolorbox}[
			title=\textbf{\footnotesize Sample Prompt: Chain-of-Programming (CoP) Applied to Quantum Computing Tasks},
			colframe=black,
			boxrule=0.5pt,
			colback=white
			]
			\footnotesize
			\textit{Prompt.} You are a quantum computing assistant. You will be presented with a programming task and a proposed solution. Your task is to determine if the provided solution is correct. If the solution is correct, confirm it with a detailed explanation. If the solution is incorrect, identify the errors and provide a corrected solution. Use the Chain-of-Programming (CoP) methodology to ensure logical progression and clarity, breaking the task into manageable steps and solving each incrementally.

			\textbf{Example:}

			\textit{Problem:} Define a simple Hamiltonian \( H = \sigma_x + \sigma_z \), where \( \sigma_x \) and \( \sigma_z \) are Pauli matrices. Find the eigenvalues and eigenvectors of \( H \).

			\textit{Proposed Solution:} The eigenvalues of \( H \) are \( 2 \) and \( -2 \).

			\textbf{Analysis:} The solution is incorrect. Using CoP methodology:

			\textbf{Step 1:} Define the Hamiltonian \( H = \sigma_x + \sigma_z \). \\
			\textbf{Step 2:} Represent the Pauli matrices:
			\[
			\sigma_x = \begin{bmatrix} 0 & 1 \\ 1 & 0 \end{bmatrix}, \quad \sigma_z = \begin{bmatrix} 1 & 0 \\ 0 & -1 \end{bmatrix}.
			\]
			\textbf{Step 3:} Add these matrices to form \( H \):
			\[
			H = \begin{bmatrix} 1 & 1 \\ 1 & -1 \end{bmatrix}.
			\]
			\textbf{Step 4:} Compute the eigenvalues and eigenvectors of \( H \) by solving \( \det(H - \lambda I) = 0 \). The eigenvalues are \( \sqrt{2} \) and \( -\sqrt{2} \). \\
			\textbf{Answer:} The correct eigenvalues are \( \sqrt{2} \) and \( -\sqrt{2} \).

			Below is the Python implementation of the task:

			\begin{minipage}{0.45\linewidth}
				\begin{lstlisting}[language=Python]
					import numpy as np
					sigma_x = np.array([[0, 1], [1, 0]])
					sigma_z = np.array([[1, 0], [0, -1]])
					H = sigma_x + sigma_z
					eigenvalues, eigenvectors = np.linalg.eigh(H)
					print("Eigenvalues: ", eigenvalues)
					print("Eigenvectors: ", eigenvectors)
				\end{lstlisting}
			\end{minipage}

			\textbf{Now analyze the following problem and solution:}

			\textit{Problem: [Insert Problem Here]} \\
			\textit{Proposed Solution: [Insert Proposed Solution Here]} \\
		\end{tcolorbox}
	\end{minipage}
	\caption{\textbf{Stage 3} Sample prompt - Enrichment using Chain-of-Programming (CoP) methodology for quantum computing problem verification.}
	\label{fig:stage3-cop}
\end{figure}

\paragraph{Stage 4: Verification with Advanced LLMs}

\begin{figure}[h!]
	\centering
	\begin{minipage}{\textwidth}
		\begin{tcolorbox}[
			title=\textbf{\footnotesize Sample Problem: Validation of Instruction-Response Pairs in Quantum Computing},
			colframe=black,
			boxrule=0.5pt,
			colback=white
			]
			\footnotesize
			\textit{Prompt.} You are an Oracle Language Model specialized in quantum computing tasks. Analyze the provided instruction-response pair and determine whether the response is correct. If the solution aligns with the instruction and is mathematically/logically valid, mark it as "Correct." Otherwise, provide a detailed justification for marking it "Incorrect." Ensure your evaluation is both precise and concise.
			\vspace{0.5\baselineskip}

			\textbf{Domain:} VALIDATION OF INSTRUCTION-RESPONSE PAIRS\par
			\vspace{\baselineskip}
		\end{tcolorbox}
	\end{minipage}
	\caption{\textbf{Stage 4:} An Oracle LLM assesses instruction-response pairs for correctness, marking a binary column in the dataset.}
	\label{fig:stage5}
\end{figure}

In this final stage, we prioritize the verification of LLM-generated solutions, aiming to ensure their accuracy and reliability. While we initially explored the possibility of incorporating advanced reasoning techniques like self-correction and self-reflection, we opted to rely on a dedicated verifier model for this task. As illustrated in (Fig.~\ref{fig:stage5}), we employ a sophisticated LLM, such as Qwen2.5-Coder-72B, functioning as a pseudo-oracle to evaluate the correctness of the generated solutions. This approach offers a robust mechanism for marking responses as either correct or incorrect without delving into iterative correction cycles.

Although self-correction methods have demonstrated potential in enhancing the stylistic and qualitative aspects of LLM outputs, their application to reasoning tasks has notable limitations. Prior studies \citep{li2023reflection,shinn2024reflexion,madaan2024self,saunders2022self,miao2023selfcheck,chen2023iterative} underscore the concept of self-correction, where LLMs iteratively refine their outputs. However, as highlighted in \citep{huang2023large}, LLMs often struggle to identify and resolve reasoning errors autonomously. Models like Reflexion \citep{shinn2024reflexion} and RCI \citep{kim2024language} depend on external feedback, such as ground-truth correctness signals, to terminate the self-correction loop effectively. Furthermore, attempts to self-correct reasoning errors can inadvertently transform accurate answers into incorrect ones, degrading overall performance \citep{huang2023large}.

Given these limitations, our method strategically leverages large-scale LLMs to serve as explicit validators. These models are tasked with identifying errors in generated solutions and marking them as correct or incorrect without attempting corrective iterations. This approach ensures a high degree of reliability in solution validation while avoiding the pitfalls associated with self-correction mechanisms in complex reasoning tasks.

\section{Conclusion}
This study centers on the exclusive \textbf{generation of synthetic data tailored for quantum computing tasks}, deliberately omitting any training or fine-tuning of large language models. No fine-tuned or trained model is provided, as our primary objective is to develop a high-quality dataset that can serve as a foundational resource for both academic and industrial research. By focusing on synthetic data generation, we aim to alleviate the significant computational demands associated with training or fine-tuning, enabling broader accessibility to advanced quantum AI research. In future work, we plan to explore the potential of training a Qwen2.5 model leveraging this dataset, paving the way for further advancements in the field.

\bibliographystyle{alphaurl}
\bibliography{iclr2025_conference}
\onecolumn
%\bibliography{quantum-template.bbl}
%\listoffigures
%\listoftables
\clearpage
\appendix

\section{Comprehensive Dataset Domains}
\label{sec:domains}
Our dataset is enriched with a very extensive set of domains in all areas of quantum physics and quantum computing, including:

\begin{enumerate}
	\item \textbf{QASM Generation:} Generate quantum circuits and produce corresponding QASM code.
	\item \textbf{Quantum Hamiltonians:} Analyze Hamiltonian time evolution and ground-state energy calculations.
	\item \textbf{Yang-Baxter Solvability:} Determine solvability of quantum models using the Yang-Baxter equation.
	\item \textbf{Trotter-Suzuki Decomposition:} Simulate Hamiltonians using Trotter-Suzuki decomposition methods.
	\item \textbf{Lindblad Dynamics:} Model open quantum systems using the Lindblad equation.
	\item \textbf{Randomized Circuits Optimization:} Optimize randomized quantum circuits to minimize error rates.
	\item \textbf{Quantum Phase Estimation:} Implement quantum phase estimation for eigenvalue calculations.
	\item \textbf{Cluster States Verification:} Prepare and verify cluster states for measurement-based quantum computation.
	\item \textbf{VQE Analysis:} Construct and optimize Variational Quantum Eigensolvers (VQE) for molecular Hamiltonians.
	\item \textbf{Quantum Algorithm Development:} Develop quantum algorithms for problems such as integer factorization and database search.
	\item \textbf{Entanglement and Quantum Information Theory:} Explore properties and applications of entangled states.
	\item \textbf{Quantum Error Correction:} Design quantum error correction codes to protect qubits from decoherence.
	\item \textbf{Semiclassical Quantum Simulation:} Simulate quantum systems with semiclassical methods.
	\item \textbf{Quantum Communication Protocols:} Develop protocols such as Quantum Key Distribution (QKD) and superdense coding.
	\item \textbf{Topological Quantum Computing:} Study fault-tolerant computing using braiding operations of anyons.
	\item \textbf{Quantum Complexity Classes:} Investigate computational problem classifications using quantum algorithms.
	\item \textbf{Quantum Thermodynamics:} Analyze thermodynamic properties of quantum systems.
	\item \textbf{Interacting Quantum Systems:} Study dynamics and correlations in interacting quantum systems.
	\item \textbf{Quantum Cryptography:} Explore quantum cryptographic protocols like QKD.
	\item \textbf{Quantum Channels:} Analyze the mathematical properties of quantum information transfer channels.
	\item \textbf{Quantum Fourier Transform:} Explore the implementation and applications of quantum Fourier transforms.
	\item \textbf{Quantum Machine Learning:} Apply quantum circuits to machine learning tasks such as classification.
	\item \textbf{Quantum State Tomography:} Reconstruct quantum states using measurement data.
	\item \textbf{Bell Inequalities and Nonlocality:} Test Bell inequalities and study quantum nonlocality.
	\item \textbf{Diagonalization of Two-Spin Hamiltonians:} Solve eigenvalue problems for two-spin systems.
	\item \textbf{Energy Eigenvalues via Perturbation Theory:} Compute eigenvalues using perturbation theory and diagonalization.
	\item \textbf{Measurement in Plus-Minus Basis:} Analyze measurement probabilities in non-standard bases.
	\item \textbf{Pauli Spin Matrices Analysis:} Explore properties and applications of Pauli matrices.
	\item \textbf{Born's Rule and State Measurement:} Apply Born's rule to calculate quantum measurement probabilities.
	\item \textbf{PennyLane Quantum Circuits:} Implement quantum circuits using the PennyLane framework.
	\item \textbf{PennyLane Circuit Analysis:} Analyze PennyLane quantum circuits for functionality and structure.
	\item \textbf{Building Molecular Hamiltonians:} Construct molecular Hamiltonians for quantum chemistry.
	\item \textbf{Variational Quantum Eigensolver (VQE):} Optimize VQE methods for specific Hamiltonians.
	\item \textbf{Subspace Search-Quantum Variational Quantum Eigensolver (SSVQE):} Find multiple eigenstates using SSVQE.
	\item \textbf{Variational Quantum State Diagonalization (VQSD):} Diagonalize density matrices with VQSD techniques.
	\item \textbf{Gibbs State Preparation:} Prepare Gibbs states for specific Hamiltonians.
	\item \textbf{The Classical Shadow of Unknown Quantum States:} Approximate quantum state properties using classical shadows.
	\item \textbf{Estimation of Quantum State Properties Based on Classical Shadows:} Estimate quantum state properties using shadow protocols.
	\item \textbf{Hamiltonian Simulation with Product Formula:} Simulate Hamiltonians using Trotter product formulas.
	\item \textbf{Simulate the Spin Dynamics on a Heisenberg Chain:} Model spin dynamics on a Heisenberg chain.
	\item \textbf{Distributed Variational Quantum Eigensolver Based on Schmidt Decomposition:} Apply distributed VQE algorithms.
	\item \textbf{Quantum Signal Processing and Quantum Singular Value Transformation:} Use quantum signal processing for operator transformations.
	\item \textbf{Hamiltonian Simulation with qDRIFT:} Simulate Hamiltonians using the qDRIFT method.
	\item \textbf{Quantum Phase Processing:} Apply quantum phase processing for signal amplification.
	\item \textbf{Variational Quantum Metrology:} Optimize quantum sensing with variational quantum metrology.
	\item \textbf{Encoding Classical Data into Quantum States:} Encode classical data into quantum states.
	\item \textbf{Quantum Classifier:} Implement quantum classifiers for machine learning.
	\item \textbf{Variational Shadow Quantum Learning (VSQL):} Approximate quantum properties with VSQL.
	\item \textbf{Quantum Kernel Methods:} Develop kernel functions for quantum machine learning.
	\item \textbf{Quantum Autoencoder:} Compress quantum data using quantum autoencoders.
	\item \textbf{Quantum GAN:} Generate quantum states using quantum GANs.
	\item \textbf{Variational Quantum Singular Value Decomposition (VQSVD):} Approximate singular value decompositions with VQSVD.
	\item \textbf{Data Encoding Analysis:} Analyze data encoding methods for quantum computation.
	\item \textbf{Quantum Neural Network Approximating Functions:} Approximate functions using quantum neural networks.
	\item \textbf{Variational Quantum Amplitude Estimation:} Estimate quantum amplitudes variationally.
	\item \textbf{Quantum Approximation Optimization Algorithm (QAOA):} Solve optimization problems with QAOA.
	\item \textbf{Solving Max-Cut Problem with QAOA:} Apply QAOA to solve Max-Cut problems.
	\item \textbf{Large-Scale QAOA via Divide-and-Conquer:} Scale QAOA for larger problem instances using divide-and-conquer techniques.
	\item \textbf{Travelling Salesman Problem:} Solve the Travelling Salesman Problem using quantum algorithms.
	\item \textbf{Jordan-Wigner Transformations:} Map spin models to fermionic systems using Jordan-Wigner transformations.
	\item \textbf{Bethe Ansatz Application:} Solve the Heisenberg spin chain spectrum using the Bethe Ansatz.
	\item \textbf{Generalized Spin Chain Compression:} Compress quantum circuits using Yang-Baxter equations in spin chain models.
	\item \textbf{Wave-Particle Duality:} Explore the dual nature of particles and waves through phenomena like the photoelectric effect.
	\item \textbf{Uncertainty Principle:} Analyze implications of Heisenberg's uncertainty principle.
	\item \textbf{Perturbation Theory:} Examine corrections to energy levels using perturbation theory.
	\item \textbf{Angular Momentum:} Investigate eigenstates and addition rules for angular momentum.
	\item \textbf{Hydrogen Atom:} Study quantization of energy levels and transitions in the hydrogen atom.
	\item \textbf{Scattering Theory:} Analyze quantum scattering and phase shifts using the Born approximation.
	\item \textbf{Quantum Tunneling:} Investigate tunneling phenomena using the WKB approximation.
	\item \textbf{Entanglement:} Examine quantum entanglement and its applications.
	\item \textbf{Time Evolution:} Analyze quantum systems' time evolution using the Schr\"odinger equation.
	\item \textbf{Quantum Measurement:} Explore wavefunction collapse and quantum measurement theory.
	\item \textbf{Quantum Harmonic Oscillator:} Study energy eigenvalues and wavefunctions of the quantum harmonic oscillator.
	\item \textbf{Spin-Orbit Coupling:} Examine spin-orbit interaction in atomic systems.
	\item \textbf{Quantum Zeno Effect:} Explore repeated measurements' effects on quantum systems.
	\item \textbf{Quantum Gates:} Construct and analyze quantum circuits with gates like Hadamard and CNOT.
	\item \textbf{Adiabatic Theorem:} Examine systems' behavior under slowly varying potentials.
	\item \textbf{Bell Inequalities:} Test Bell inequalities and analyze their implications.
	\item \textbf{Superposition Principle:} Investigate the principle of superposition in quantum mechanics.
	\item \textbf{Quantum Decoherence:} Analyze the loss of coherence in quantum systems.
	\item \textbf{Topological Quantum States:} Explore the robustness of topological quantum states.
	\item \textbf{Quantum Cryptography:} Study principles and protocols of quantum cryptography.
	\item \textbf{Quantum Eraser:} Examine implications of the quantum eraser experiment.
	\item \textbf{Quantum Teleportation:} Demonstrate quantum teleportation principles.
	\item \textbf{Path Integral Formulation:} Utilize the path integral approach in quantum mechanics.
	\item \textbf{Quantum Annealing:} Analyze quantum annealing for optimization problems.
	\item \textbf{Berry Phase:} Study geometric phase in quantum systems.
	\item \textbf{Quantum Cloning:} Explore the no-cloning theorem in quantum mechanics.
	\item \textbf{Density Matrix Formalism:} Describe mixed states using density matrices.
	\item \textbf{Quantum Computation:} Explore basic concepts of quantum computation.
	\item \textbf{Relativistic Quantum Mechanics:} Solve problems involving the Klein-Gordon or Dirac equations.
	\item \textbf{Quantum Field Theory:} Introduce concepts of quantum field theory.
\end{enumerate}

This comprehensive set of domains offers a foundational resource for advancing quantum algorithms, problem-solving, machine learning, and optimization techniques in quantum physics and quantum computing.

\section{Performance Evaluation}

To assess the efficiency of the QuantumLLMInstruct pipeline, we conducted a performance evaluation on two different setups: a high-performance HuggingFace A100G machine and a local NVIDIA 4080-powered system. The evaluation measured the execution times for generating quantum problem instructions using the Qwen2.5-Coder-1.5B-Instruct model and its smaller variant, Qwen-7B Coder. The results provide insights into the computational demands and scalability of the pipeline across various quantum domains.

\paragraph{Execution Times on HuggingFace A100G Machine}
The execution times recorded on the HuggingFace A100G machine showcase the model's efficiency in generating problem instructions in real-time for fundamental quantum domains.

\begin{table}[ht]
	\centering
	\begin{tabular}{|l|c|c|}
		\hline
		\textbf{Domain}                    & \textbf{Problem ID} & \textbf{Time Taken (seconds/problem)} \\ \hline
		Berry Phase                        & 1, 3                & 3.23, 2.78                            \\ \hline
		Wave-Particle Duality              & 2                   & 2.20                                   \\ \hline
		Quantum Classifier                 & 4, 5                & 2.78, 2.64                            \\ \hline
	\end{tabular}
	\caption{Performance Metrics for Generating Quantum Problem Instructions (Model: Qwen2.5-Coder-1.5B-Instruct on HuggingFace A100G Machine)}
	\label{tab:instruction_generation_time}
\end{table}

These results highlight the ability of the model to rapidly generate high-quality problem instructions, with each problem taking less than 4 seconds on average.
\paragraph{Execution Times on a Local Machine (NVIDIA 4080)}
The local system, powered by an NVIDIA 4080 GPU, demonstrated the pipeline's adaptability to more common hardware configurations. However, the execution times varied significantly depending on the complexity of the problem and the availability of required templates.
\begin{table}[h!]
	\centering
	\begin{tabular}{|l|c|c|}
		\hline
		\textbf{Domain}                                           & \textbf{Problem ID} & \textbf{Time Taken (seconds/problem)} \\ \hline
		Quantum Probability Calculations                         & 6, 11               & 230.14, 186.72                       \\ \hline
		Measurement in Qubit Basis                               & 7                   & Error ('template')                    \\ \hline
		Estimation of Quantum State Properties                  & 9                   & 114.30                                \\ \hline
		Bell Inequalities and Nonlocality                        & 10                  & 145.84                                \\ \hline
		Entanglement and Quantum Information Theory             & 11                  & 186.72                                \\ \hline
		Quantum Error Correction                                 & 12                  & Error ('template')                    \\ \hline
		PennyLane Quantum Circuits                               & 13                  & 154.16                                \\ \hline
	\end{tabular}
	\caption{Time Taken to Generate Problems for Each Quantum Domain (Model: Qwen-7B Coder on NVIDIA 4080)}
	\label{tab:problem_generation_time}
\end{table}

\end{document}